# Neuromorphic computing with nanoscale spintronic oscillators

Jacob Torrejon[1*], Mathieu Riou[1*], Flavio Abreu Araujo[1*], Sumito Tsunegi[2], Guru Khalsa[3†], Damien Querlioz[4], Paolo Bortolotti[1], Vincent Cros[1], Akio Fukushima[2], Hitoshi Kubota[2], Shinji Yuasa[2], M. D. Stiles[3] and Julie Grollier[1]

1 - Unité Mixte de Physique, CNRS, Thales, Univ. Paris-Sud, Université Paris-Saclay, 91767 Palaiseau, France

2 - National Institute of Advanced Industrial Science and Technology (AIST), Spintronics Research Center, Tsukuba, Ibaraki 305-8568, Japan

3 - Center for Nanoscale Science and Technology, National Institute of Standards and Technology, Gaithersburg, Maryland 20899-6202, USA

4 - Centre de Nanosciences et de Nanotechnologies, CNRS, Univ. Paris-Sud, Université Paris-Saclay, 91405 Orsay France

* jacob.torrejondiaz@cnrs-thales.fr, mathieu.riou@cnrs-thales.fr, flavio.abreuaraujo@cnrs-thales.fr

† now at: Cornell University, Department of Materials Science and Engineering, Ithaca, NY 14853-1501, USA

**Neurons in the brain behave as non-linear oscillators, which develop rhythmic activity and interact to process information[1]. Taking inspiration from this behavior to realize high density, low power neuromorphic computing will require huge numbers of nanoscale non-linear oscillators. Indeed, a simple estimation indicates that, in order to fit a hundred million oscillators organized in a two-dimensional array inside a chip the size of a thumb, their lateral dimensions must be smaller than one micrometer. However, despite multiple theoretical proposals[2–5], there is no proof of concept today of neuromorphic computing with nano-oscillators. Indeed, nanoscale devices tend to be noisy and to lack the stability required to process data in a reliable way. Here, we show experimentally that a nanoscale spintronic oscillator[6–8] can achieve spoken digit recognition with accuracies similar to state of the art neural networks. We pinpoint the regime of magnetization dynamics leading to highest performance. These results, combined with the exceptional ability of these spintronic oscillators to interact together, their long lifetime, and low energy consumption, open the path to fast, parallel, on-chip computation based on networks of oscillators.**

Spintronic nano-oscillators, illustrated in Fig. 1a, are nanopillars composed of two ferromagnetic layers separated by a non-magnetic spacer. Charge currents become spin-polarized when they flow through these junctions and generate torques on the magnetizations[9,10] leading to sustained magnetization precession at frequencies between hundreds of megahertz to several tens of gigahertz. Magnetization oscillations are converted into voltage oscillations through magneto-resistance. The resulting radiofrequency oscillations, up to tens of millivolts[11], can be detected by measuring the voltage across the junction (Fig. 1b). Spin-torque nano-oscillators are therefore simple and ultra-compact: their lateral size can be scaled down to ten nanometers and their power consumption reduced down to one microwatt[12]. As they have the same structure as current magnetic memory cells, they are compatible with Complementary Metal Oxide Semiconductor technology, have high endurance, operate at room temperature, and can be fabricated in large numbers (up to hundreds of millions today) on a chip[13]. Finally, just as the frequency of a neuron is modified by the spikes received from other neurons, the frequencies of spin-torque nano-oscillators are highly sensitive to the magnetization dynamics of neighboring oscillators to which they are coupled[14,15]. Together, these features of spin-torque nano-oscillators are promising for neuromorphic computing with large arrays of coupled oscillators[16–21]. However, their use for an actual computing task has never been physically demonstrated.

Our idea consists in exploiting the amplitude dynamics of spin-torque nano-oscillators for neuromorphic computing. Indeed, their oscillation amplitude $\tilde{V}$ (dotted blue line in Fig. 1b) is robust to noise, due to the confinement provided by the counteracting torques exerted by the injected current and magnetic damping[22]. In addition, $\tilde{V}$ is highly non-linear as a function of the injected current and intrinsically depends on past inputs[14]. Exploiting the amplitude dynamics of spin-torque nano-oscillators therefore gathers in one single nanodevice the two most crucial properties of neurons: non-linearity and memory, which would otherwise require the combination of several lumped components and a much larger area on chip using conventional electronics[23]. To compute we encode neural inputs in the current injected in the oscillator $I(t)$ and use the amplitude response $\tilde{V}(t)$ as the neural output.

Our nanoscale oscillators are circular magnetic tunnel junctions, with a 6 nm thick FeB free layer 375 nm in diameter, which have magnetic vortex ground states (see Methods). We measure directly the signal amplitude dynamics $\tilde{V}(t)$ with a microwave diode. Fig. 1c shows the non-linear response of the amplitude $\tilde{V}$ to a dc current $I$: $\tilde{V} \propto \sqrt{(I - I_{th})}$, where $I_{th}$ is the current threshold for steady oscillations to occur[14]. Using an arbitrary waveform generator, we inject a varying current though the junctions in addition to the dc current, using the set-up schematized in Fig. 1d. The resulting voltage oscillations, recorded with an oscilloscope, are shown in Fig. 1e. The amplitude of the oscillator varies in response to the injected dc current, with a relaxation time inducing a few hundred nanoseconds memory of past inputs[22].

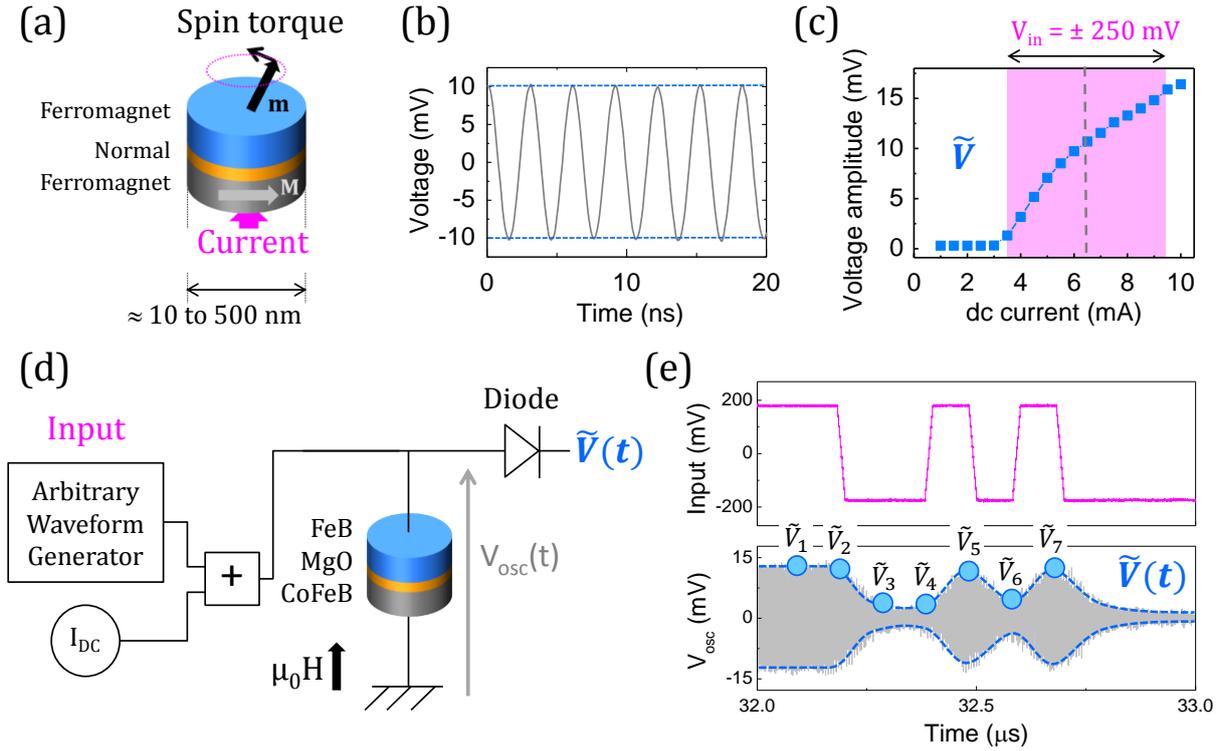

*Figure 1: Spin-torque nano-oscillator for neuromorphic computing. (a) Schematic of a spin-torque nano-oscillator, consisting of a non-magnetic spacer between two ferromagnetic layers. (b) Measured ac voltage emitted by the oscillator as a function of time $V_{osc} = \tilde{V}(t)\cos(\omega t + \varphi)$, for a steady current injection of 7 mA at the external magnetic field $\mu_0 H$ = 430 mT. The dotted blue line highlights the amplitude $\tilde{V}$. (c) ac voltage amplitude $\tilde{V}$ as a function of current at $\mu_0 H$ = 430 mT. The typical resulting excursion of voltage amplitude is highlighted in magenta when an input signal $V_{in}$ of ±250 mV is injected (here for a bias current of 6.5 mA and $\mu_0 H$ = 430 mT.) (d) Schematic of the experimental set-up. A dc current $I_{DC}$ as well as a fast-varying waveform encoding the input are injected in the spin-torque nano-oscillator. The microwave voltage $V_{osc}$ emitted by the oscillator in response to the excitation is measured with an oscilloscope. For computing, the amplitude $\tilde{V}$ of the oscillator is used, and measured directly with a microwave diode. (e) Input (magenta) and measured microwave voltage $V_{osc}$ emitted by the oscillator as a function of time. Here $I_{DC}$ = 6 mA, $\mu_0 H$ = 430 mT. The envelope $\tilde{V}$ of the oscillator signal is highlighted in blue.*

Recent studies have elegantly pointed out that time-multiplexing can be leveraged to use a single oscillator to emulate a full neural network[24–26]. Here we use this approach, a form of "reservoir computing"[4,5] (see Methods), to demonstrate the ability of spin-torque nano-oscillators to realize neuromorphic tasks. We perform experimentally a benchmark task of spoken digits recognition. The input data, taken from the TI-46 database[27], are audio waveforms of isolated spoken digits (0 to 9) pronounced by five different female speakers (Fig. 2a). The goal is to recognize the digits, independent of the speaker.

Neural networks classify information through chain reactions: neuron after neuron, each input undergoes a series of non-linear transformations[28]. In a trained network, the same digit always triggers a similar chain reaction even if it is pronounced by different speakers, whereas different digits generate different chain reactions, thus allowing pattern recognition. An input can trigger a chain reaction in space by using ensembles of neurons: the state of downstream neurons depends on the state of upstream neurons. But an input can also trigger a chain reaction in time by constantly exciting a single non-linear oscillator with memory: the state of the oscillator in the future depends on the state of the oscillator in the past. Here, we use this approach, which simplifies the hardware as only one oscillator is needed, but requires preprocessing the input: each point of the audio file is converted to a fast binary sequence designed to generate a chain reaction of oscillator amplitude variations[24].

The procedure is illustrated in Fig. 2a-d and detailed in Methods. As acoustic features are mainly encoded in frequencies[29], we filter each audio file to $N_f$ different frequency channels (a standard procedure in speech recognition), which are then concatenated in intervals of duration $\tau$, as displayed in Fig. 2b. For preprocessing, each of these segments is multiplied by a randomly filled binary matrix (of dimensions $N_f \times N_\theta$). In this way, each point of the input audio file is converted in a binary sequence of duration $\tau$ composed of $N_\theta$ points separated by a time step $\theta$ ($\tau = N_\theta \theta$). When this preprocessed input (Fig. 2c) is applied as a current to our spin-torque oscillator, the resulting amplitude variations $\tilde{V}(t)$ (Fig. 2d) function as a set of $N_\theta$ neurons coupled in time (we take $N_\theta$ samples $\tilde{V}_i$ per interval $\tau$). For spoken digit recognition, we emulate $N_\theta$ = 400 neurons, and use $\theta$ = 100 ns (about one fifth of the oscillators' relaxation time) to set the oscillator in transient state.

The responses of the oscillator's voltage amplitude $\tilde{V}(t)$ are recorded for each utterance of each spoken digit. The goal of the subsequent training process, done on a computer, is to choose a linear combination of these responses (e.g. sets of $\tilde{V}_i$) for each digit such that the sum is one for that digit and zero for the rest (see Methods). As each digit has been pronounced ten times by each of the five speakers, we can use part of the data to determine the coefficients (training), and the rest to evaluate the recognition performance (testing) (see Methods). In order to assess the impact of our oscillator on the quality of recognition, we always perform a control trial without the oscillator. In that case, the pre-processed input traces are used directly to reconstruct the outputs on the computer, without going through the experimental set-up.

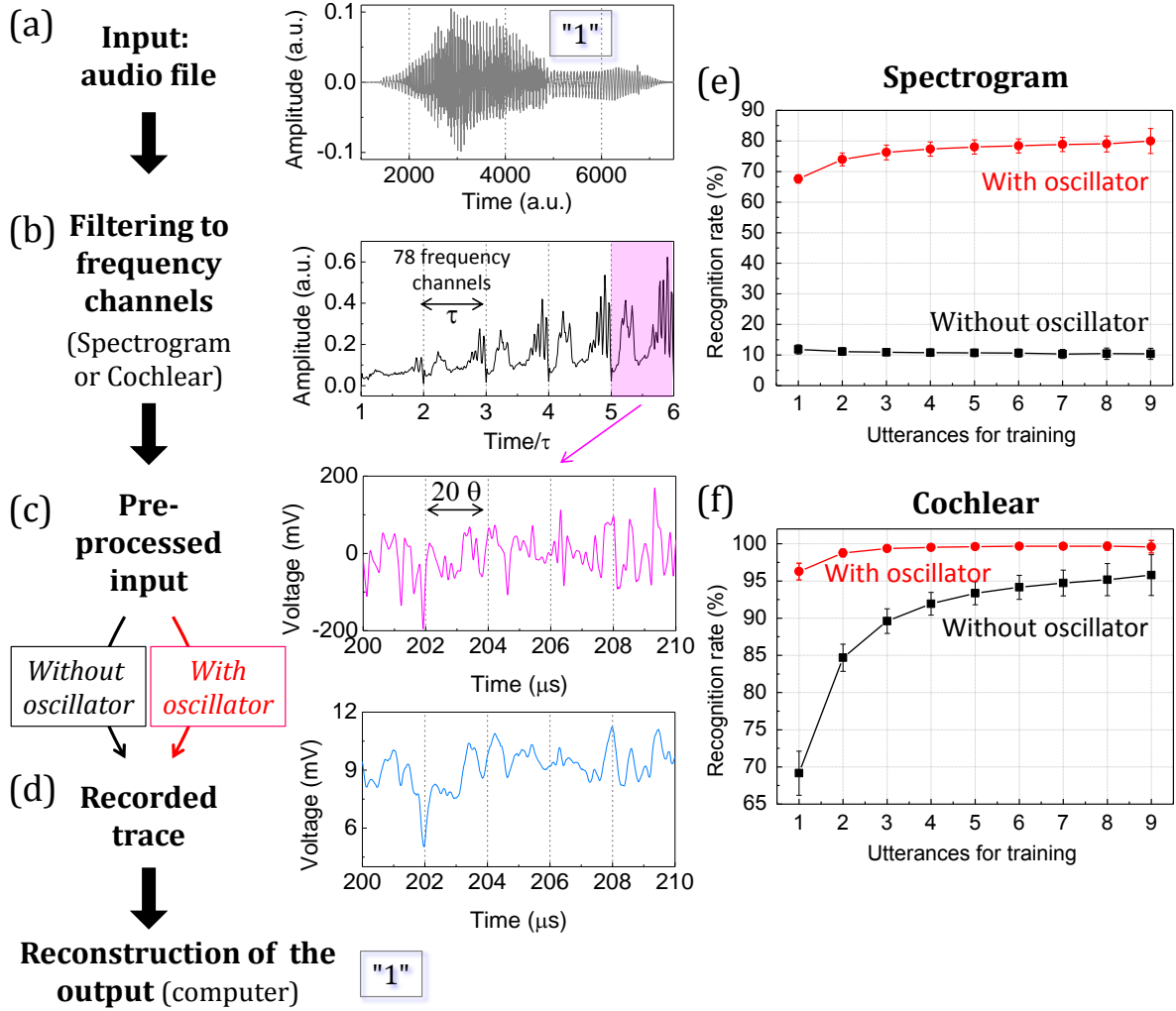

*Figure 2: Spoken digit recognition (a-d) Principle of the experiment. (a) Audio waveform corresponding to the digit 1 pronounced by speaker 1. (b) Filtering to frequency channels for acoustic feature extraction. The cochlear model filters each point of the audio waveform in 78 frequency channels (65 in the case of the spectrogram model). The frequency channels are then concatenated in intervals of duration $\tau$ to form the filtered input. (c) Pre-processed input (zoom). The filtered input is multiplied by a randomly filled binary matrix (masking process), resulting in 400 points separated by a time step $\theta$ of 100 ns in each interval of duration $\tau$ ($\tau = 400\ \theta$). (d) Oscillator output. Envelope $\tilde{V}(t)$ of the experimental oscillator's emitted voltage amplitude ($\mu_0 H$ = 430 mT, $I_{DC}$ = 6 mA). The 400 values of $\tilde{V}(t)$ per interval $\tau$ ($\tilde{V}_i$, sampled with a time step $\theta$) emulate 400 neurons. **(e-f)** Spoken digit recognition rates in the testing set as a function of the number of utterances N used for training (since there are many ways to pick the N utterances used for training, the recognition rate is an average over all 10!/(10-N)!N! combinations of N utterances out of the 10 in the data set) **(e)** for the spectrogram filtering ($\mu_0 H$ = 430 mT, $I_{DC}$ = 6 mA) and **(f)** for the cochlear filtering ($\mu_0 H$ = 448 mT, $I_{DC}$ = 7 mA). The red curves are the experimental results using the magnetic oscillator. The black curves are control trials, in which the pre-processed inputs are directly used for reconstructing the output on a computer, without going through the experimental set-up. The error bars correspond to the standard deviation of the word recognition rate based on training with all possible combinations.*

In both Fig. 2e and f, the improvement shown in the experimental results over the control results indicates that the spintronic oscillator significantly improves the quality of spoken digit recognition, despite the added noise concomitant to its nanometric size. In Fig. 2e (linear spectrogram filtering), we present a case in which the acoustic feature extraction, achieved by Fourier transforming the audio waveform over finite time windows, plays a minimal role in classification. Without the oscillator (black), the recognition rates are consistent with random choices. With the oscillator (red line in Fig. 2e), the recognition rate is improved by +70 % and reaches values up to 80 %. This example highlights the crucial role played by the oscillator in the recognition process. Using a non-linear cochlear filtering[30] (Fig. 2f), which is the standard in reservoir computing[24–26], and has been optimized based on the behavior of biological ears, we achieve recognition rates up to 99.6 %, as high as the state of the art. Compared to the control trial, the oscillator reduces the error rate by a factor of up to 15. Our results with a nanoscale spintronic oscillator are therefore comparable to the recognition rates obtained with more complicated electronic or optical systems (between 95.7 % and 99.8 % for the same task with cochlear filtering)[23–26,29].

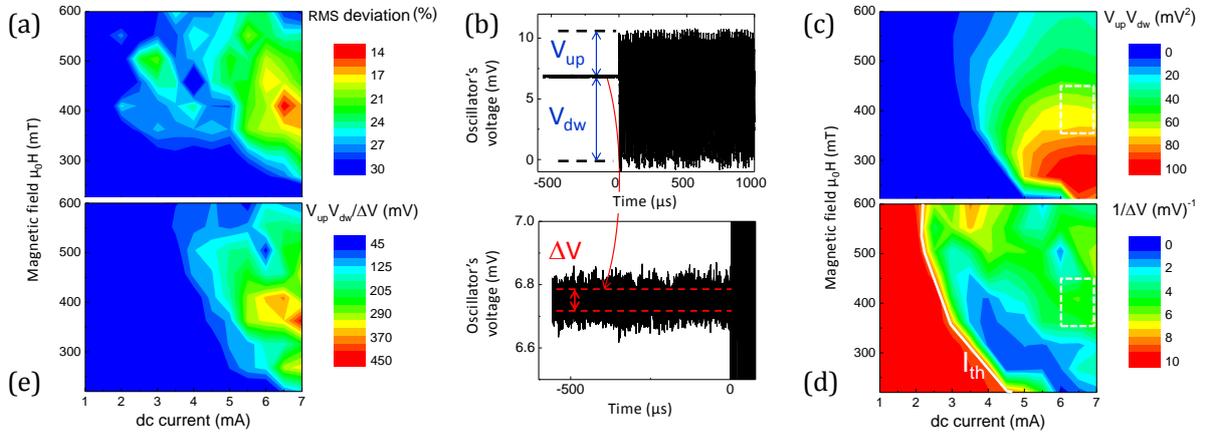

*Figure 3: Conditions for optimal waveform classification and identification of important oscillator properties.* *The task consists of recognizing sine from square waveforms with the same period. The target for the output reconstructed from the oscillator's response is one for square, zero for sine. We emulate 24 neurons $\tilde{V}_i$, $\tau = 24\,\theta$. **(a)** Root mean square of output-to-target deviations: map as a function of dc current $I_{DC}$ and magnetic field $\mu_0 H$. **(b)** Extraction of parameters from the time traces of the oscillator's response. **Top:** Maximum positive ($V_{up}$) and negative ($V_{dw}$) variations of the oscillator's amplitude in response to the varying preprocessed input. **Bottom:** Noise $\Delta V$ of the oscillator's voltage at steady state under $I_{DC}$. **(c)** Maximal response of the oscillator to the input $V_{up} \times V_{dw}$: map in the $I_{DC}$ - $\mu_0 H$ plane. **(d)** Inverse of amplitude noise $1/\Delta V$: map in the $I_{DC}$ - $\mu_0 H$ plane. The threshold current $I_{th}$ is indicated by a white solid line. In (c) and (d), the optimal bias condition range for waveform classification is marked by a white dashed rectangle (currents between 6 mA to 7 mA and magnetic fields from 350 mT to 450 mT). **(e)** Map of the maximal amplitudes over noise ratio $V_{up} \times V_{dw} / \Delta V$ showing that these parameters largely determine the performance of the oscillator.*

The optimal operating conditions for pattern recognition with our spin-torque nano-oscillator are determined by the oscillation amplitude and noise. We use a simpler task, classification of sine and square waveforms with the same period[28], to investigate the ability of the oscillator to classify waveforms in a wide range of injected dc currents $I_{DC}$ and applied magnetic fields $\mu_0 H$ (see Methods). As can be seen in Fig.3a, the quality of pattern recognition, characterized by the root mean square of deviations between reconstructed output and target, varies from 10 % to over 30 % depending on the bias conditions. The oscillator performs well when it responds strongly to the time-varying preprocessed input, with large amplitude variations $V_{up}$ and $V_{dw}$ for both positive and negative directions (Fig.3b, top). On the other hand, it performs poorly when the oscillator's noise $\Delta V$, (the standard deviation of the voltage amplitude noise) is high (Fig.3b, bottom). As shown in Fig.3b, we extract these parameters from the time traces of the oscillator's emitted voltage at each bias point, and plot the corresponding figures of merit $V_{up} \times V_{dw}$ (Fig.3c) and $1/\Delta V$ (Fig.3d) as a function of current and field. The red regions of large oscillation amplitude in Fig.3c correspond to low magnetic fields where the magnetization is weakly confined and high currents where the spin torque on magnetization is maximal. The blue regions of high noise in Fig.3d correspond to areas just above the threshold current $I_{th}$ for oscillation where the oscillation amplitude $\tilde{V}$ is growing rapidly as a function of current and is becoming sensitive to external fluctuations[14]. As can be seen by comparing Fig.3c and d, a range of bias conditions highlighted by a dotted white box (currents between 6 mA to 7 mA and magnetic fields from 350 mT to 450 mT) features wide variations of oscillation amplitudes and low noise. In this region, root mean square deviations below 15 % are achieved, for which there is no classification errors between sines and squares. The good match between the map of $V_{up} \times V_{dw}/\Delta V$ in Fig.3e and the map giving the performance of classification (Fig.3a) confirms that the best conditions for classification correspond to regions of optimal compromise between low noise and large amplitude variations. The necessity of high signal-to-noise ratio for efficient neuromorphic computing highlighted here for magnetic oscillators is a general guideline that applies to any type of nanoscale oscillator.

As a conclusion, our pattern recognition results show that simple, ultra-compact spintronic oscillators have all the properties needed to emulate collections of neurons: non-linearity, memory and stability. The additional ability of these oscillators to connect through the currents and magnetic fields they emit opens the path to large scale hardware neural networks exploiting magnetization dynamics for computing[14–21].

**Methods**

**Samples**

Magnetic tunnel junctions (MTJs) films with a stacking structure of buffer/PtMn(15)/ $Co_{71}Fe_{29}$(2.5)/Ru(0.9)/ $Co_{60}Fe_{20}B_{20}$(1.6)/$Co_{70}Fe_{30}$(0.8)/ MgO(1)/ $Fe_{80}B_{20}$(6)/ MgO(1) /Ta(8)/Ru(7) (thicknesses in nm) were prepared by ultra-high vacuum (UHV) magnetron sputtering. After annealing at 360 °C for 1 h, the resistance-area products (RA) were ≈ 3.6 $\Omega\mu m^2$. Circular-shape MTJs with a diameter ≈ 375 nm were patterned using Ar ion etching and e-beam lithography. The resistance of the samples is close to 40 $\Omega$, and the magneto-resistance ratio is about 135 % at room temperature. The FeB layer presents a vortex structure as the ground state for the dimensions used here. In a small region called the core of the vortex, the magnetization spirals out of plane. Under dc current injection, the core of the vortex steadily gyrates around the center of the dot with a frequency in the range of 250 MHz to 400 MHz for the oscillators we consider here. Vortex dynamics driven by spin-torque are well-understood, well-controlled and have been shown to be particularly stable[22].

**Measurement set-up**

The experimental implementation for spoken digit recognition and sine/square classification tasks is illustrated in Fig. 1d. The preprocessed input signal $V_{in}$ is generated by a high frequency arbitrary waveform generator and injected as a current through the magnetic nano-oscillator. The sampling rate of the source is set to 200 MHz (20 points per interval of time $\theta$) for the spoken digit recognition task, and 500 MHz (50 points per interval of time $\theta$) for the classification of sines and squares. The peak to peak variation of the input signal is 500 mV, which corresponds to peak to peak current variations of 6 mA, as illustrated in Fig.1c (part of the incoming signal is reflected due to impedance mismatch). The bias conditions of the oscillator are set by a dc current source and an electromagnet which applies a field perpendicular to the plane of the magnetic layers. The oscillating voltage emitted by the nano-oscillator is rectified by a planar tunnel microwave diode, of bandwidth from 0.1 GHz to 12.4 GHz and response time of 5 ns. The input dynamic range of the diode is between 1 µW and 3.15 mW corresponding to a dc output level between 0 mV and 400 mV, respectively. We use an amplifier to accommodate the emitted power of the nano-oscillator to the working range of diode. The output signal is then recorded by a real time oscilloscope. In Figs. 1b, c and e, 2d and 3b c, d and e the amplitude of the signal emitted by the oscillator is shown without amplification (the signal measured after the diode has been divided by the total amplification of the circuit, $\approx$ +21 dB). If, due to sampling errors, the measured oscillators' envelope is shifted with respect to the input, classification accuracy can be degraded. We use alignment marks to align our measurements with the input when we reconstruct the output. The alignment precision is ± 1 ns.

**Reservoir Computing: general concepts**

In machine learning, a reservoir is a network of recurrently and randomly connected non-linear nodes[4,5]. When an input signal is injected in the reservoir, it is mapped to a higher dimensional space in which it can become linearly separable. The key insight behind reservoir computing is that the network does not need any tuning: all connections inside the reservoir are kept fixed. Only external connections (between the reservoir and an output layer) are trained to achieve the desired task.

In other words, reservoir computing requires the generation of complex non-linear dynamics but as a trade-off, learning is greatly simplified. For efficient reservoir computing, several requirements related to dynamical properties of the network should be satisfied. First, different inputs should be trigger different dynamics (separation property) while similar inputs should generate similar dynamics (approximation property) allowing efficient classification. Second, the reservoir state should not only depend on present inputs but also on recent past inputs. This short-term memory, called fading memory, is essential for processing temporal sequences for which the history of the signal is important.

A single non-linear oscillator can emulate a reservoir when it is set in transient dynamics by a fast varying input[24]. The loss of parallelism is compensated by an additional pre-processing input step: the input is multiplied by a fast varying mask, which allows defining virtual nodes interconnected in time through the resultant oscillator dynamics. This approach provides a drastic simplification of the reservoir scheme for hardware implementations, and has been realized in hardware with optical or electronic oscillators assembled from several components[23–26].

**Spoken digit recognition**

For this task, the inputs are taken from the NIST TI-46 data corpus[30]. The input consists of isolated spoken digits said by five different female speakers. Each speaker pronounces each digit 10 times. The 500 audio waveforms are sampled at a rate of 12.5 kHz and have variable time lengths.

We have used two different filtering methods: spectrogram and cochlear model. Both filters break the word in several time intervals $N_\tau$ of duration $\tau$ and analyze the frequency content in each interval $\tau$ either through Fourier transform (spectrogram, 65 channels, $N_\tau$ = 24 to 67, Fig. 2b) or a more complicated non-linear approach (cochlear, 78 channels, $N_\tau$ = 14 to 41). The input for each word is composed of an amplitude for each of $N_f$ = 65 (78) frequency channels times $N_\tau$ time intervals. This input is preprocessed by multiplying the frequency content for each time interval by a mask matrix containing $N_f$ x $N_\theta$ random binary values giving at total of $N_\tau$ x $N_\theta$ values as input to the oscillator (Fig. 2c). Here, we are modelling $N_\theta$ = 400 input neurons, each of which is connected to all of the frequency channels for each time interval.

Each input value is consecutively applied to the oscillator as a constant current for a $\theta \approx$ 100 ns time interval, which is about 5 times smaller than the relaxation time of the oscillator, as recommended in reference[24]. This time is short enough to guarantee that the oscillator is maintained in its transient regime so the emulated neurons are connected to each other, but it is also long enough to let the oscillator respond to the input excitation. The amplitude of the ac voltage across the oscillator is recorded for off-line post-processing (Fig. 2d).

The post-processing of the output consists of two distinct steps. The first is called the training (or learning) process and the second is called the classification (or recognition) process. The goal of training is to determine a set of weights $w_{i,\theta}$ where *i* indexes the desired digit. These weights are used to multiply the output voltages to give 10 x $N_\tau$ output values, which are then averaged over the $N_\tau$ time intervals to give 10 output values, $y_i$, which should ideally be equal to the target values $\tilde{y}_i =$ 1.0 for the appropriate digit and 0.0 for the rest. In the training process, a fraction of the utterances are used to train these weights and in the classification process, the rest of the utterances are used to test the results.

The optimum weights are found by minimizing the difference between $\tilde{y}_i$ and $y_i$ for all the words used in the training. In practice, optimal values are determined by using techniques for extracting meaningful eigenvalues from singular matrices such as the linear Moore-Penrose pseudo-inverse operator (denoted by a dagger symbol †). If we consider the matrix target $\tilde{Y}$ containing the targets $\tilde{y}_i$ for all the time steps τ used for the training and $S$ the matrix containing all the neuron responses for all the time steps τ used for the training, then the matrix W containing the optimal weights is given by: $W = \tilde{Y}S^\dagger$. This step is done on a computer and takes several seconds. In the future, real time processing at the nanosecond time scale can be realized using fully parallel networks of interacting nano-oscillators.

During the classification phase, the 10 reconstructed outputs corresponding to one digit are averaged over all the time steps τ of the signal and the digit is identified by taking the maximum value of the 10 averaged reconstructed outputs. The averaged reconstructed output corresponding to the digit should be close to one and the others should be close to zero. The efficiency of the recognition is evaluated with the word success rate, which is the rate of digits correctly identified. The training can be done using more or less data (here utterances). We always trained the system using the 10 digits spoken by the 5 speakers. The only parameter changed is the number of utterances used for the training. If we use *N* utterances for training, we use the remaining 10-*N* other utterances for testing. However, some utterances are very well pronounced while other are hardly distinguishable. As a consequence, the resulting recognition rate depends on which *N* utterances are picked for training in the set of 10 (e.g, if *N* = 2, the utterances picked for training could be the first and second, but also second and third, or sixth and tenth, or any other of the 10!/(8! 2!) combinations of 2 picked out of 10). In order to avoid this bias, the recognition rates that we give in the paper are the average of the results over all possible combinations. The error bars corresponds to the standard deviation of the word recognition rate. The raw spectrogram is not complex enough to allow a correct reconstruction of the target during the training. Adding the oscillator brings complexity and suppresses this phenomenon.

**Sine and squares classification**

For this classification task the input is a random sequence of 160 sines and squares with the same period: the first half of the sequence for training and the second half for classification. Each period is discretized into 8 points separated by a time step τ. The preprocessing consists in multiplying the value of each point by the same binary sequence generated by a random distribution of +1 and -1 values. In comparison with spoken digit recognition, the mask is a binary vector instead of a binary matrix. The fast binary sequence contains 24 values, so during each time step τ, 24 neurons $\tilde{V}_i$ are emulated.

The target $\tilde{y}$ for the network output $y$ is 0 for all the trajectories in response to a sine and 1 for all the trajectories in response to a square. The best weights are found by linear regression as was previously explained for spoken digit recognition task. For sine/square recognition, we record five points instead of one for each "neuron" when we measure the output of the oscillator. During post-processing, we use these additional states between $\tilde{V}_i$ and $\tilde{V}_{i+1}$ to increase the number of coefficients available for solving the problem, and thus increase classification accuracy. In addition, the best performance does not necessarily correspond to a target in exact phase with the oscillator's output. The standard deviation of the RMS value of $(V_{output} - V_{target})$, obtained with 10 repetitions, is around 1 %.

**Acknowledgements**

This work was supported by the European Research Council ERC under Grant bio*SPIN*spired 682955. The authors would like to thank Laurent Larger, Bogdan Penkovsky and François Duport for useful discussions.

**Author contributions**

The study was designed by JG and MDS, samples were optimized and fabricated by ST, experiments were performed by JT and MR, numerical studies were realized by FAA, MR and GK, all authors contributed to the analysis of results and writing of paper.

**Additional information**

Reprints and permissions information is available online at www.nature.com/reprints. Correspondence and requests for materials should be addressed to J.G.

**Competing financial interests**

The authors declare no competing financial interests.